\input{aipcheck}

\documentclass[
    ,final            
  ]
  {aipproc}
\layoutstyle{6x9}
\usepackage{wrapfig}
\usepackage{caption}[2008/04/01]

\begin{document}

\title{Jet quenching measurements with ATLAS at LHC}


\author{W. K. Brooks, for the ATLAS Collaboration}{
  address={Institute for Advanced Studies in Physics and Engineering\\
  Valpara\'iso Center for Science and Technology\\
  Universidad T\'ecnica Federico Santa Mar\'ia\\
  Valpara\'iso, Chile}
}

\begin{abstract}
	A broad program of measurements is planned for heavy ion collisions in ATLAS. With up to a factor of 30 increase in collision energy compared to existing data, significant new insights are anticipated to be obtained with the first data measured. Global features of the LHC collisions will be accessible with the early data and will set the stage for the precision measurements to follow. ATLAS is particularly well suited for exploration of "jet quenching," the extinction of energetic jets in the hot dense medium. Observations of heavy quark jet suppression will be possible with unprecedented energy reach and statistical precision, potentially yielding new insights into the basic mechanisms involved.
\end{abstract}

\maketitle


\section{Context}

Ultrarelativistic heavy ion collisions explore extreme conditions of strongly interacting matter. These studies connect to a diverse set of forefront topics ranging from saturation of gluonic fields to string theory \cite{Kharzeev2010}. Experiments at the Brookhaven National Laboratory RHIC facility indicate that a new state of matter is formed in A-A collisions with $\sqrt{s}=$ 200~GeV/nucleon based on a number of key observations. This picture is consistent with extensive lattice QCD studies predicting a phase transition as well as the results from CERN experiments at lower energies suggesting the onset of a phase transition. Among the key observations, the discovery of {\em{jet quenching}} is perhaps the clearest indication of the formation of an ultra-dense system - the "disappearance" of one of the two jets in back-to-back jet production in central collisions, together with the associated suppression of high-$p_T$ jet events.

Further exploration of this phenomenon, as well as systematic investigations of these collisions, will be possible at the LHC. With a planned ultimate energy of 5.5 TeV/nucleon for Pb-Pb collisions, the anticipated program benefits from a factor of 30 increase in $\sqrt{s}$ as well as from highly hermetic detector systems with essentially complete coverage in precision calorimetry. The complementary heavy ion studies planned for the ATLAS, CMS, and ALICE experiments at LHC together cover the full spectrum in $p_T$ and pseudorapidity $\eta$ needed to understand the complex and potentially surprising collision events that are anticipated. In the following is a brief discussion of the planned program for the ATLAS experiment, with an overview of the detector and the heavy ion program followed by the prospects for advances in jet quenching studies.

\section{Overview of ATLAS}

ATLAS and CMS are the two broad-focus experiments at LHC, with complementary designs that provide excellent resolution and highly hermetic coverage in calorimetry, tracking, and muon systems. The ATLAS calorimetry is a particular strong point for jet quenching studies, with an angular coverage down to $|\eta|\approx$ 5 (0.001$\rm^o$) and fine-grained segmentation. 

ATLAS is comprised of three major detector component types: the inner tracking systems, the electromagnetic and hadronic calorimeters, and the muon systems. Each is made up of several subsystems; for an in-depth discussion of these as well as of the superconducting magnets, trigger, readout, data acquisition, controls and other system components, see \cite{ATLASPAPER,ATLASPERFORMANCE}. The inner tracking systems, which operate in a 2 T solenoidal magnetic field, consist of three subsystems with a total of more than 86 million electronic channels. The calorimeters, which represent nearly 190,000 channels, are configured in four subsystems, three of which are based on liquid argon technology. The muon systems, which have in total one million channels, make use of toroidal magnetic fields of up to 2.5 and 3.5 T in the barrel and endcap region, respectively. An average of 99\% of the channels in each of these subsystems were operational well in advance of the 2009 data taking. A complex, three-level trigger and data acquisition system selects and transports the data to disk; calibration, reconstruction, and simulation software are exercised on the worldwide computing grid of LHC, a network of $\rm10^5$ computers located in more than 50 countries around the world.

\section{The ATLAS heavy ion program}
The expectations for the heavy ion data are based on extrapolations from RHIC energies and from the first p-p data acquired at LHC in 2009 and 2010. Because the energy available in the collisions is so much greater than for previous measurements, the first order of business is to check the validity of these extrapolations \cite{BARBARA,PETER}. Of particular interest are the energy density and degree of collectivity of the response, as expressed by global observables such as the charged particle multiplicity and the elliptic flow \cite{PETER2}. These observables are critical for validating the physical picture extrapolated from RHIC measurements, and they can begin to be determined with a relatively small integrated luminosity from the first few days of heavy ion collision data. Subsequent topics that can be addressed with a one-month run at the planned luminosity of $\rm10^{27}cm^{-2}s^{-1}$ include studies of quarkonia production and of hard probes such as jets and high-$\rm p_T$ hadrons. In addition to light quark jets, heavy quark jets and heavy mesons can be accessed.

The expected sensitivity of ATLAS to heavy ion events has been fully explored using the 
HIJING \cite{HIJING} event generator and the full simulation and reconstruction chain of ATLAS. Some highlights of these studies are mentioned here, and the results are more fully documented elsewhere \cite{PPR}. The first variable to be determined from the data is the event centrality, on which most other observables depend. The event centrality can be estimated from the response of the calorimeters; according to the simulations, the number of collisions within the event, the number of collision participants, and the impact parameter are all monotonically related to the total energy deposits in the forward calorimeter, the electromagnetic calorimeters, and the hadronic calorimeters. The estimation of the event centrality using such a primitive measure from several calibrated detector systems will facilitate consistency checks over a wide range in pseudorapidity $\eta$.

\begin{wrapfigure}{R}{0.5\textwidth}
    \includegraphics[width=0.5\textwidth]{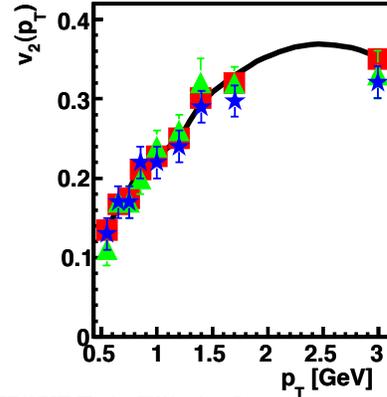}
    \vspace{-0.07\textwidth}
    \captionsetup{labelformat=simple,labelsep=period,name=FIGURE,margin=10pt,font=footnotesize,labelfont=bf}
    \caption{Elliptic flow reconstructed from ATLAS simulated data with an impact parameter of 10.7 fm using three different methods.}
\end{wrapfigure}

A fundamental observable depending on the event centrality is the charged particle multiplicity $dN_{ch}/d\eta$. It has been shown that $dN_{ch}/d\eta$ can be independently estimated by counting the number of pixel hits in the first, second, and third layers of the ATLAS pixel 
detector \cite{ADAM}. The method requires application of a simple correction function that is determined primarily from geometry. Other methods have also been explored, such as the "tracklet" method [8] that combines an event vertex with hits in the first two pixel layers. The determination of $dN_{ch}/d\eta$ will allow a first estimate of the energy density and will eliminate significant uncertainties in extrapolations of this quantity from lower energy collisions.

The distribution of the transverse energy with angle $dE_T/d\eta$ has been shown to be accurately determined in ATLAS using a simple energy sum over the electromagnetic and hadronic calorimeter cells. According to simulations, this method reproduces the event-by-event distribution in $E_T$ to the few percent level over the wide angular range $|\eta|~\leq 5$.

Elliptic flow $v_2$ is a parameter characterizing the azimuthal angular distribution of particles emitted with respect to the reaction plane defined by the impact parameter vector and the beam axis. It is defined through the Fourier expansion of this angular distribution 

\vspace{0.005\textheight}
\begin{center}
$dN/d(\phi - \Psi) = N_0(1+2v_1cos(\phi-\Psi) + 2v_2 cos(2(\phi-\Psi))+...)$
\end{center}
\vspace{0.005\textheight}

\noindent where $\Psi$ is the angle of the reaction plane within ATLAS and $\phi$ is the azimuthal angle of an individual particle.
Measurement of elliptic flow in ATLAS has been explored using three different methods, the reaction plane method, the two-particle correlation method, and the Lee-Yang Zeros method \cite{PPR}. HIJING events have been simulated for impact parameters b=2.3, 7.0, and 10.7 fm, and the three methods give results that are consistent within approximately 5\% for the most central collisions and somewhat better for the most peripheral collisions as seen in Fig. 1. The resolution for determination of the reaction plane is particularly good for ATLAS since this can be performed independently in several different detectors systems, all of which have complete coverage in azimuthal angle $\phi$. In the simulation studies, the resolution correction averaged over seven subsystems for central events was 0.86, very close to unity, demonstrating the excellent event plane resolution available in ATLAS.

The capability for heavy quarkonium reconstruction has been studied in the dimuon channel. As an example, the mass resolution found for the $\Upsilon$ 1S state was found to be approximately 120 MeV, easily adequate for separation of the 1S and 2S $\Upsilon$ states \cite{JIRI}. This capability will significantly extend the reach of the color screening studies with quarkonia that have been performed previously at RHIC and CERN. Additional studies in the dimuon channel have been performed to determine the ATLAS performance for identification of $J/\Psi$ and $Z$ bosons in the heavy ion environment. In these studies, a mass resolution for $Z$ bosons of approximately 2.2 GeV was found, comparable to that found in the p-p environment, with a background level of less than 10\% for central collisions.

\begin{center}
\begin{figure}
  \includegraphics[height=.35\textheight,angle=90]{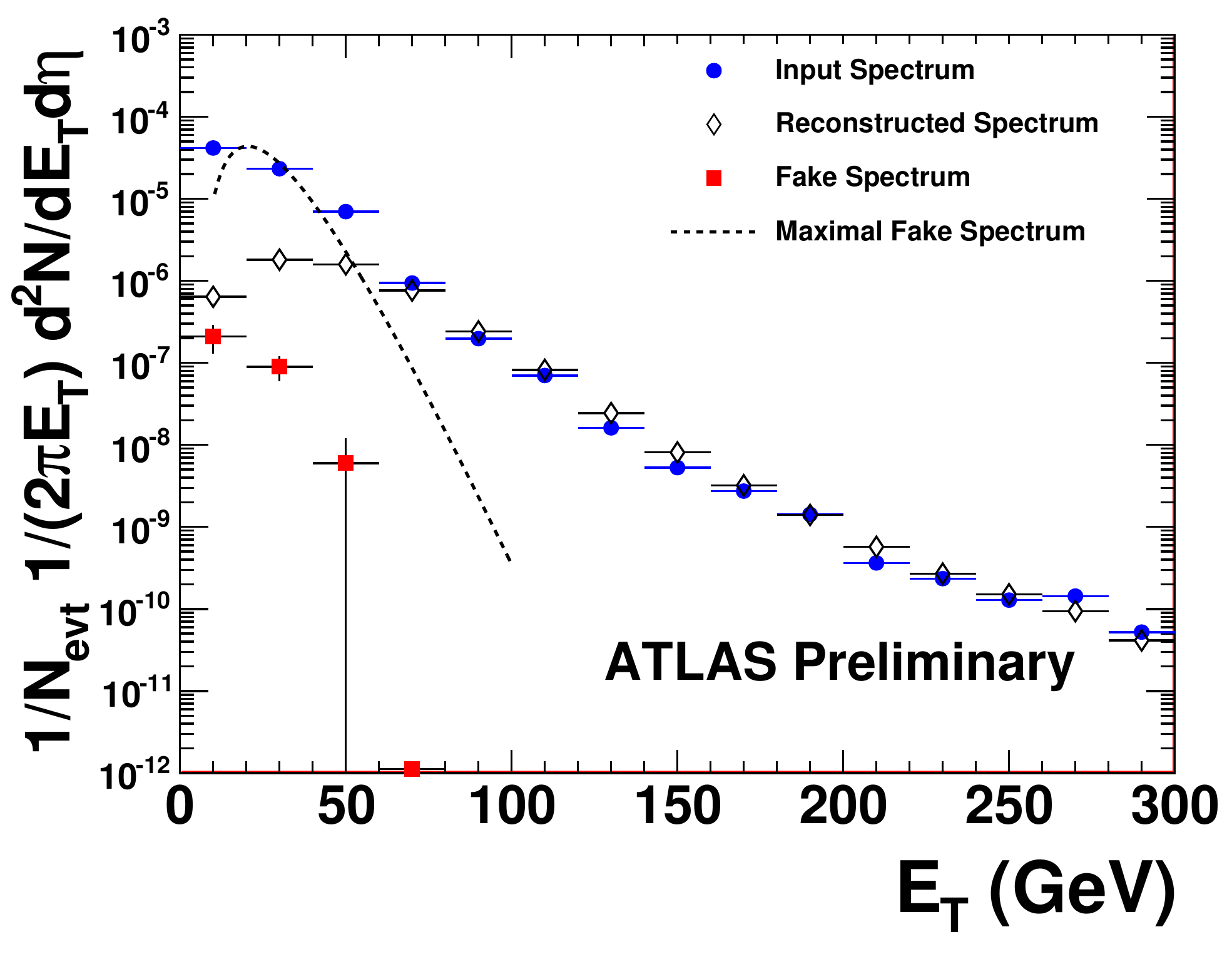}
  \includegraphics[height=.27\textheight]{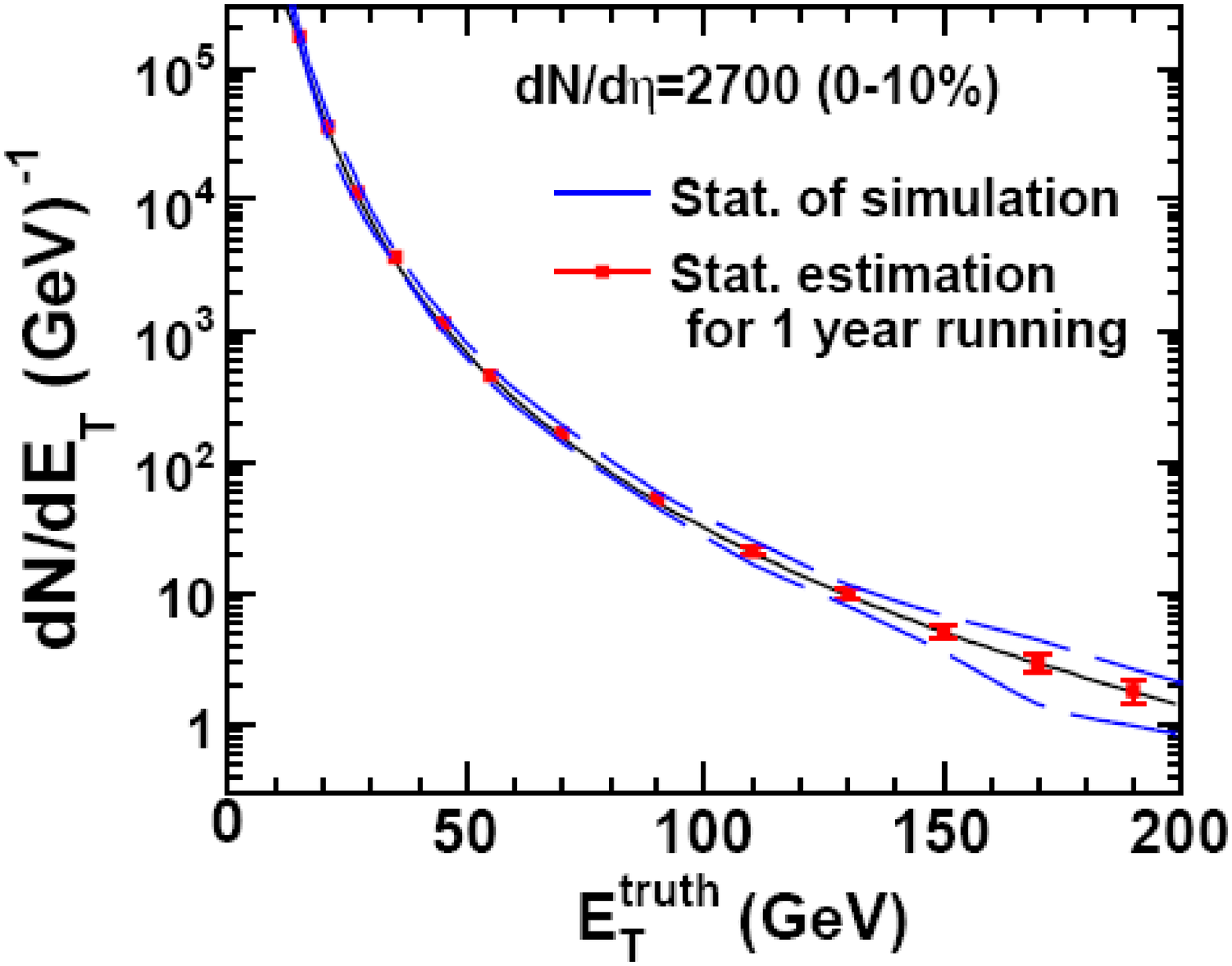}
  \caption{(left) Jet reconstruction for simulated ATLAS heavy ion events. The dashed line represents the absolute fake jet rate from pure HIJING events prior to background jet rejection. Red squares show residual fake jets after the rejection. Input and raw reconstructed spectra are shown in circles and diamonds respectively. (right) Direct photon reconstruction for simulated ATLAS heavy ion events assuming an integrated luminosity of $\rm0.5~nb^{-1}$.}
\end{figure}
\end{center}

\vspace{-0.05\textheight}

\section{Jet suppression physics}
Jet suppression is an important component of the ATLAS heavy ion physics program, and an area where the excellent calorimetry and hermetic coverage of the calorimeters is very beneficial. Simulations have been performed for light quark jets and photon-tagged jets as well as for heavy quark signatures. The background from the low-energy QCD interactions is significant, thus, methods have been developed for identifying light quark jets using a background subtraction procedure, and to identify photons with good efficiency while rejecting signals from other neutral particles \cite{PPR}.

In the heavy ion environment, jets in ATLAS after background subtraction are reconstructed with essentially full efficiency above 70-80 GeV, with no additional corrections required \cite{NATHAN} as seen in Fig. 2 (left panel). Further, according to simulations, there is essentially no distortion of the extracted light quark jet fragmentation function above $z>0.1$, and the distortions due to the heavy ion background for $0.003<z<0.1$ can be corrected reliably \cite{SPOUSTA2009}. Extensive studies of direct photons and photon-jet correlations have been performed with HIJING simulated data. In order to reject background, isolation cuts and shower shape cuts have been developed which can be optimized to preserve good reconstruction efficiency while rejecting hits from neutral hadrons. These studies show that excellent direct photon statistics will be obtained as seen in Fig. 2 (right panel). Further, photon-jet correlations are clearly accessible  as seen in Fig. 3 (left panel) \cite{BAKER}.

\begin{center}
\begin{figure}
\includegraphics[height=.25\textheight]{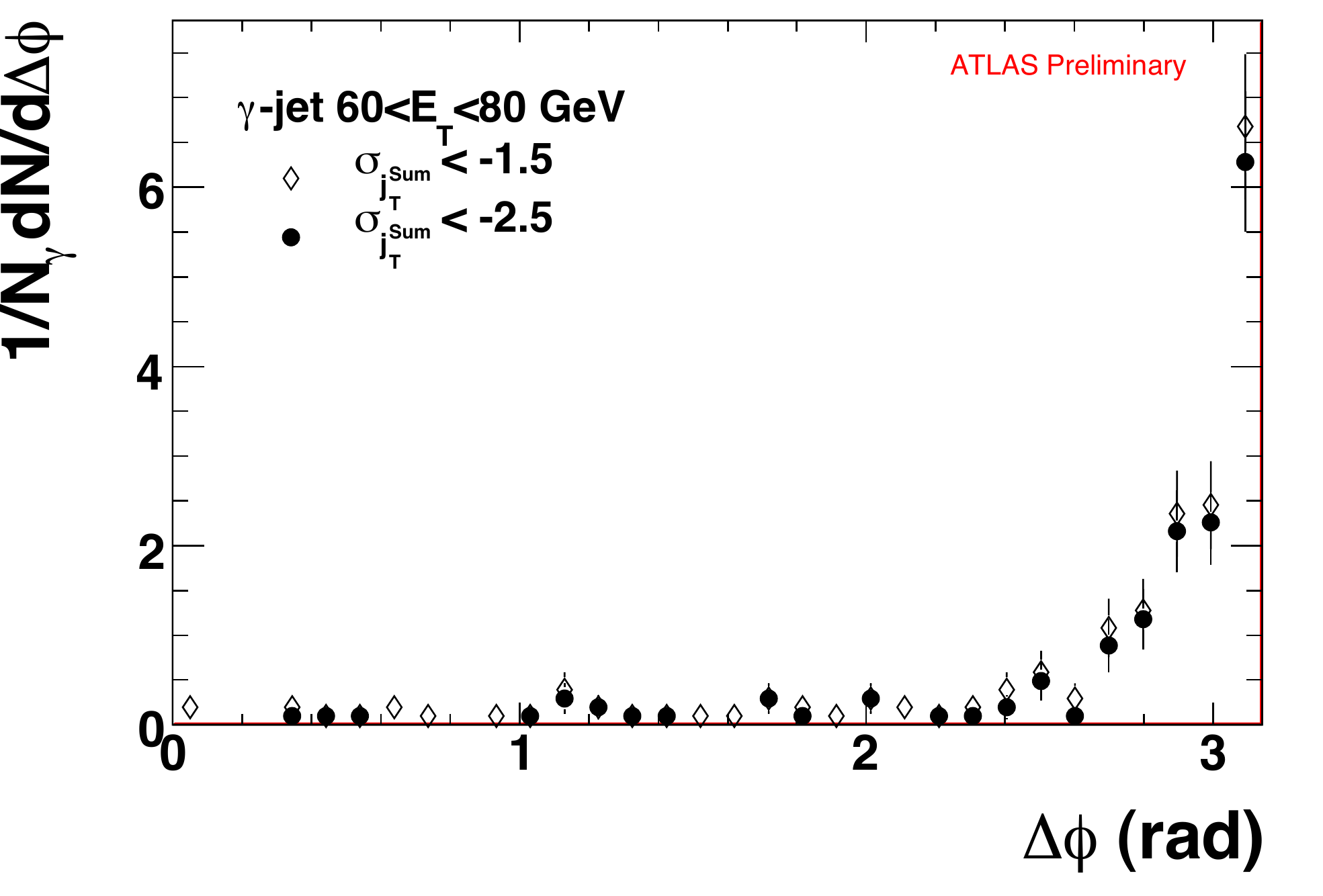}
\includegraphics[height=.25\textheight]{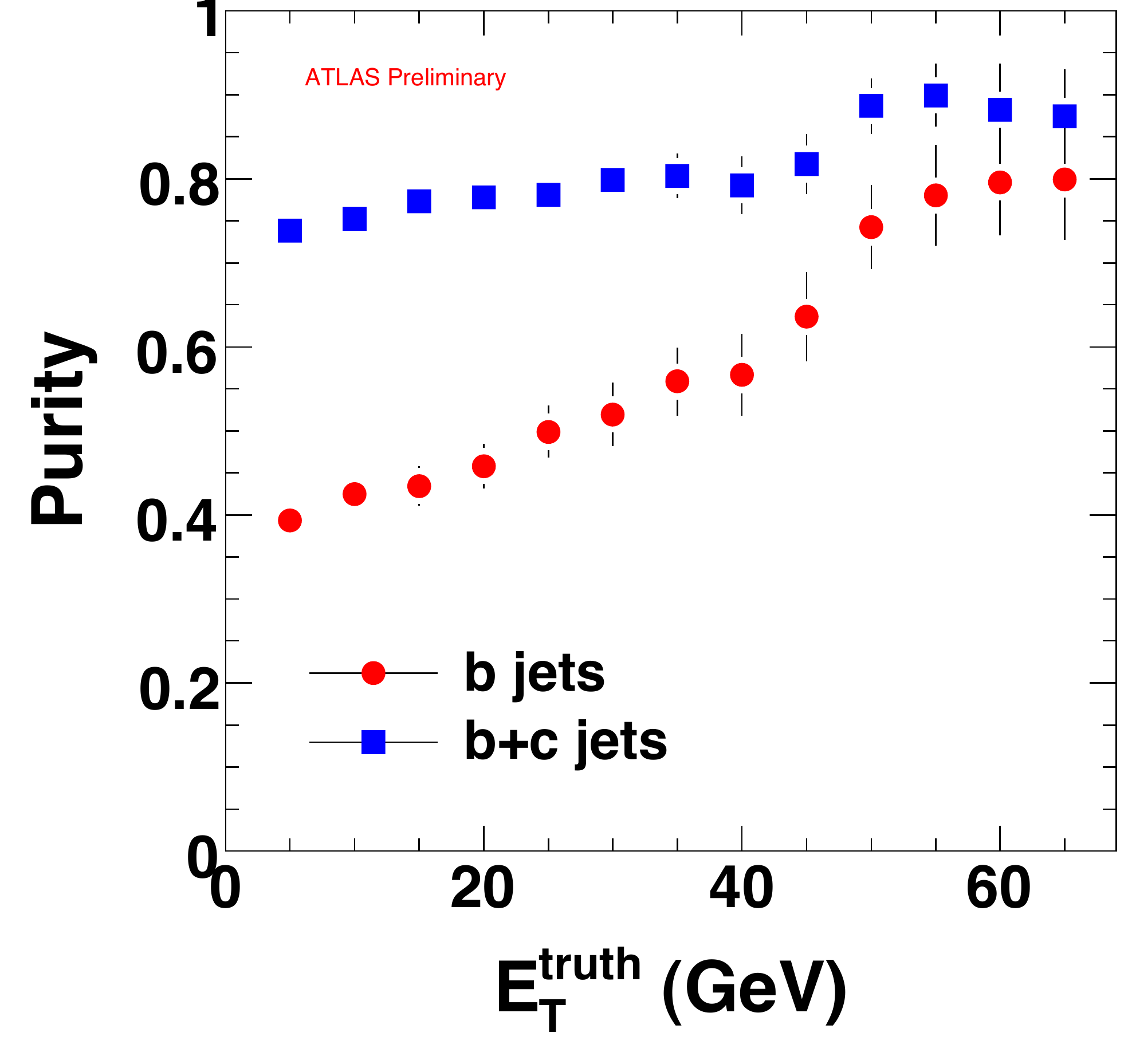}
\caption{(left) Jet-photon correlations from ATLAS heavy-ion simulations as a function of azimuthal angle difference, for tight and loose quality cuts \cite{PPR}. (right) Purity of the heavy quark sample in detected dimuons from semileptonic decays in ATLAS heavy-ion simulations. }
\end{figure}
\end{center}

\vspace{-0.045\textheight}
Heavy quark jets have been studied in the dimuon channel, using the semileptonic decay of b and c quarks. According to these simulations, the purity of the dimuon sample above 50 GeV is substantial as can be seen in Fig. 3 (right panel). Heavy quark suppression at RHIC remains an unsolved puzzle \cite{KOLIU2007}, although it can be explained in alternative pictures \cite{BORIS}, and thus it is of high scientific interest to address it at the LHC. Further studies relevant to heavy quark jet suppression in ATLAS are planned, e.g., in the semileptonic decay to $\rm e^+$ and $\rm e^-$, and in reconstruction of heavy mesons using detached vertex tagging.

\vspace{-0.03\textheight}

\begin{theacknowledgments}
\vspace{-0.02\textheight}
  The author acknowledges partial support under Chilean FONDECYT grant 1080564.
\end{theacknowledgments}



\bibliographystyle{aipproc}   


\vspace{-0.01\textheight}

\end{document}